\documentclass[journal=accacs,manuscript=article]{achemso}

\usepackage{mhchem}
\usepackage[T1]{fontenc}

\graphicspath{{Figures/}}

\author{Hassan Ouhbi}
\author{Ulrich Aschauer}
\email{ulrich.aschauer@dcb.unibe.ch}
\affiliation[University of Bern]
{Department of Chemistry and Biochemistry, University of Bern, Bern, Switzerland}

\title{Water oxidation catalysis on reconstructed NaTaO$_3$ (001) surfaces}

\begin{document}


\begin{abstract}
Polar perovskite oxide surfaces are subject to structural reconstruction as a possible stabilisation mechanisms, which changes the surface structure and hence the surface chemistry. To investigate this effect, we study the oxygen evolution reaction (OER) on the reconstructed (001) surface of NaTaO$_3$, by means of density functional theory (DFT) calculations and compare it with the non-polar (113) surface of the same material. For the clean surface the reconstruction has a beneficial effect on the catalytic activity, lowering the minimal overpotential from 0.88 V to 0.70 V while also changing the most active reaction site from Na to Ta. Under photocatalytic conditions, the Ta sites are covered by oxygen adsorbates, rendering a lattice oxygen site on the NaO terrace the most active with a very low overpotential of 0.32 V. An alternative surface reconstruction stable in contact with water leads to an oxygen coupling mechanism with an overpotential of 0.52 V. Our results show that terraced surface reconstructions enable novel reaction pathways with low overpotentials that do not exist on other non-polar NaTaO$_3$ surfaces nor on non-polar surfaces of chemically similar materials such as SrTaO$_2$N.
\end{abstract}

\section{Introduction}

\begin{figure}
	\centering
	\includegraphics{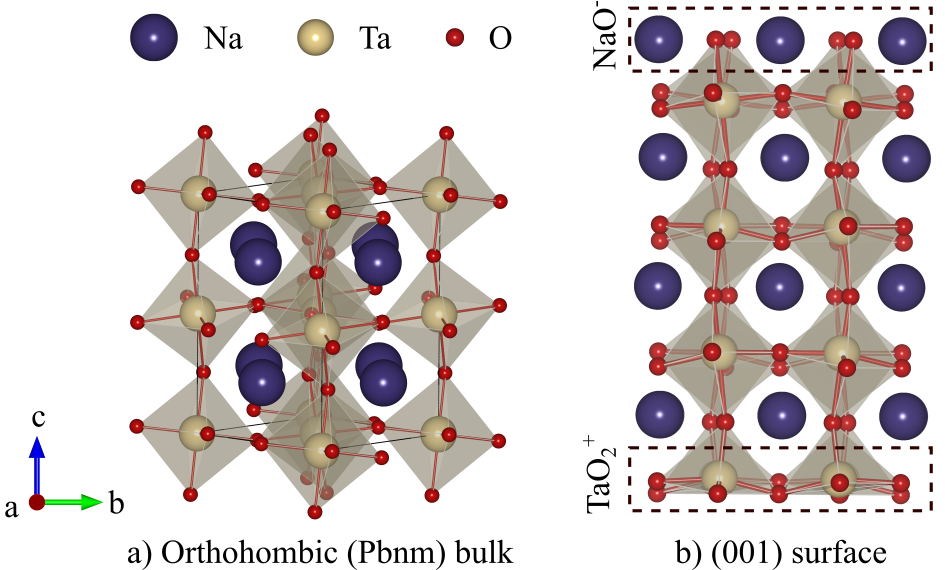}
	\caption{NaTaO$_3$ a) bulk and b) non-reconstructed (001) surface.}
	\label{fig:bulk_surface_structure}
\end{figure}

Among perovskite structured tantalates, sodium tantalate NaTaO$_3$ has received much attention for photocatalytic water splitting due to its outstanding quantum and photocatalytic efficiency \cite{Grabowska2016, Modak2016}. In this material the larger A-site cation (Na) is 12-fold coordinated by oxygen and located at the corners of a pseudocubic unit cell, while the center of this cell is occupied by the smaller B-site cation (Ta) that is octahedrally coordinated by oxygen (see Figure \ref{fig:bulk_surface_structure}a). For NaTaO$_3$, the formal charges of Na, Ta, and O are +1, +5, -2 respectively. Consequently, the (001) surface is formed by stacking of charged TaO$_2$ and NaO layers, with formal charges of +1 and -1 respectively (see Figure \ref{fig:bulk_surface_structure}b). The surface is hence of type III in Tasker's classification \cite{Tasker1979} and is polar and unstable due to its diverging electrostatic energy \cite{Noguera2000}. Such an instability can be compensated either by adsorption of charged species or electronic or ionic reconstruction \cite{Noguera2013, Deacon-Smith2014a}.

While the atomic structure of NaTaO$_3$ surfaces has to the best of our knowledge not been investigated experimentally and only sparsely by theory \cite{Liu2015a, Zhao2018bc}, the closely related compound KTaO$_3$ has been investigated much more intensively. Secondary ion mass spectrometry revealed a gradual K enrichment of cleaved (001) surfaces that was interpreted as a bulk to surface migration of K atoms \cite{Szot2000}. A similar evolution of the surface was observed by helium atom diffraction, where an almost equal proportion of TaO$_2$ and KO terminations right after cleaving gradually converted to a pure KO termination \cite{Li2001}. Supported by density functional theory (DFT) calculations \cite{Fritsch1999}, the authors postulate a terraced surface structure and a migration of K atoms to KO step edges that leads to the disappearance of the TaO$_2$ termination \cite{Li2001, Li2003}. An ionic reconstruction with alternating KO and TaO$_2$ rows is also supported by more recent DFT calculations \cite{Deacon-Smith2014}, while other calculations reported a purely electronic reconstruction and a transition from a KO to a TaO$_2$ terminated surface when going from oxygen-poor to oxygen-rich conditions \cite{Wang2018}. The terraced surface structure of KTaO$_3$ is supported by a recent scanning tunnelling and atomic force microscopy study combined with DFT calculations \cite{Setvin2018}. While the cleaved surface exposes TaO$_2$ and KO terminated terraces that follow the domain pattern of the incipient ferroelectricity in this material, annealing ultimately converts it into a labyrinth-like structure of alternating TaO$_2$ and KO terminated terraces running along the [110] direction. The surface energy was determined to be minimal for terrace widths of 4-5 unit cells. After exposure to liquid water, the same authors found another structure type consisting of [001] oriented alternating single unit-cell rows of KO termination hat are stabilized by dissociative water adsorption \cite{Setvin2018}, which was confirmed by another theoretical study \cite{Zhao2018bc}.

Given the similarity in ionic and electronic structure of NaTaO$_3$ and KTaO$_3$, it seems sensible that NaTaO$_3$ will assume a surface reconstruction similar to KTaO$_3$ to compensate polarity. Significantly lower surface energies compared to the non-reconstructed terminations were reported for the terraced reconstructed surfaces of both NaTaO$_3$ and KTaO$_3$ with terrace widths of 3 cells \cite{Zhao2018bc}. However for both materials a so-called cation-exchange reconstruction was reported to be even lower in energy, the formation of which may however be kinetically hindered \cite{Zhao2018bc} leading to the experimentally reported prevalence of the terraced reconstruction for KTaO$_3$ \cite{Setvin2018}. Due to the larger driving force, the appearance of the cation-exchange reconstruction is more likely for NaTaO$_3$ than KTaO$_3$ \cite{Zhao2018bc} but due to a lack of experimental data is it yet unclear if the kinetic barriers retain the terraced reconstruction also for the former material. For this reason we work here under the assumption of a terraced reconstruction also for NaTaO$_3$.

Surface features such as the step edges appearing on this reconstruction are known to affect the surface chemistry of oxides. As such it was recently shown that stepped RuO$_2$ (110) surfaces are less active for the oxygen evolution reaction (OER) compared to flat oxygen-terminated surfaces \cite{Dickens2017}. For hematite Fe$_2$O$_3$ on the other hand, the surface orientation was shown to have a much stronger effect on the OER overpotential than reaction sites at different positions with respect to step edges \cite{Zhang2016b}. It is thus unclear what effect the terraced surface structure of NaTaO$_3$ will have on the photocatalytic OER activity. For this reason, we compare, by means of DFT calculations, the OER activity of the reconstructed terraced (001) surface with the one of the non-polar (113) surface already investigated in our previous work \cite{Ouhbi2018}. For the clean surface, stable at low potentials, we find Ta sites at the center of terrace on the reconstructed (001) surface to be the most active. At higher potentials relevant for photocatalytic applications, the surfaces are covered with O adsorbates and we find a lattice-oxygen mechanism at the center of the NaO terraces to be most active with a very low overpotential of 0.32 V. These results show that the reconstructed surface structure enables reaction pathways that are not favorable on alternative non-polar surface orientations nor on non-polar surfaces of chemically similar materials such as SrTaO$_2$N.

\section{Methods}

Density functional theory calculations were performed with the Quantum ESPRESSO package \cite{Giannozzi2009}, using the Perdew-Burke-Ernzerhof (PBE) exchange correlation functional \cite{Perdew1996}. Ultrasoft pseudopotentials \cite{Vanderbilt1990} with Na(2s, 2p, 3s), Ta(5s, 5p, 5d, 6s) and O(2s, 2p) valence states were used to describe electron-nuclear interactions and wave functions were expanded in plane waves up to a kinetic energy cutoff of 40 Ry combined with 320 Ry for the augmented density.

Surface slabs were constructed based on the experimental orthorhombic NaTaO$_3$ \textit{Pbnm} perovskite structure \cite{Kennedy2006} that was fully relaxed at the PBE level of theory. The reconstructed (001) surface with terraces of width $w$=4 running along the [110] direction was modelled with an 8 atomic layer thick slab (4 NaO and 4 TaO$_2$) with lateral dimensions 5.49$\times$22.20 \AA, from which half of the topmost NaO layer has been moved to the bottom surface of the slab. Periodic images of the slab are separated by a vacuum of 15 \AA\ along the surface normal direction and the bottom two atomic layers as well as the moved NaO half layer were fixed at bulk positions. In addition, a dipole correction \cite{Bengtsson1999} was applied along the z-direction for all slab calculations. Reciprocal-space integration was performed with a 4$\times$1$\times$1 Monkhorst-Pack mesh \cite{Pack1977}. Structures were relaxed with a threshold of 10$^{-3}$ eV$\AA^{-1}$ and 10$^{-6}$ eV for forces and total energies respectively.

We characterise the stability of a surface by its surface energy, which for our stoichiometric slabs is given by
\begin{equation}
	\gamma = \frac{E_{slab} - N \times E_{bulk}}{2A},
\end{equation}
where $E_{slab}$ is the total energy of the slab, $N$ is the number NaTaO$_3$ bulk formula units in the slab, $E_{bulk}$ is the energy per formula unit in the bulk and $A$ is the surface area of the slab.

All reactions energies $\Delta G(U_b, \mathrm{pH})$ under given pH and potential ($U_b$) conditions were calculated within the computational standard hydrogen electrode approach \cite{Norskov2004}:
\begin{equation}
\Delta G(U_b, \mathrm{pH}) = \Delta E + (\Delta\mathrm{ZPE} - T\Delta S) - \mathrm{e}U_b - k_BT\cdot \ln(10)\cdot \mathrm{pH}
\label{eq:deltaG}
\end{equation}
where $\Delta E$ is the DFT-calculated reaction energy, -e$U_b$ and $-k_B T \cdot \ln(10) \cdot \mathrm{pH}$ account for the electron and proton reservoirs respectively and $\Delta$ZPE and $\Delta S$ are the changes in zero point energy and entropy respectively of the reaction step as given in our previous paper \cite{Ouhbi2018}. We characterize the OER activity of a given surface by its thermodynamic overpotential $\eta$, which is the potential for which all PCET steps have free-energy changes smaller than the equilibrium potential of 1.23 V:
\begin{equation}
	\eta = \frac{\max(\Delta G)}{e} - 1.23 \mathrm{V}
\label{eq:OP}
\end{equation}

\section{Results and discussion}

\subsection{Surface reconstruction}
\begin{figure}
	\centering
	\includegraphics{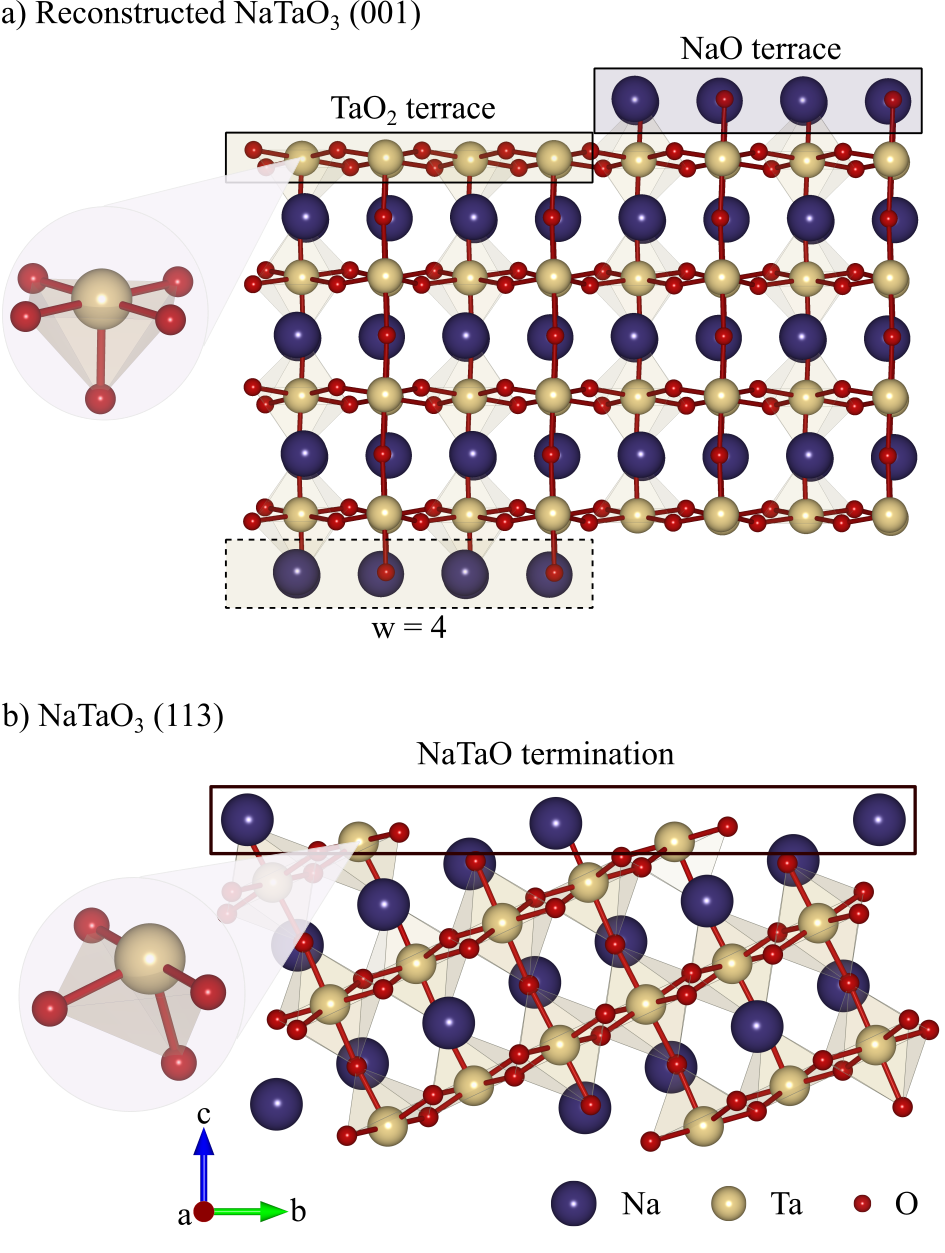}
	\caption{Slab models of a) the reconstructed (001) and b) the (113) surface of NaTaO$_3$. The surface termination layers are highlighted by solid boxes and the $w=4$ unit-cell wide NaO half layer moved to the bottom surface is shown highlighted by the dashed box. The insets show the differently coordinates Ta sites on the (001) and (113) surface.}
	\label{fig:slab_models}
\end{figure}

For the reconstructed NaTaO$_3$ (001) surface with a terrace width $w$=4 shown in Figure \ref{fig:slab_models}a), we predict a surface energy of 1.20 Jm$^2$ (see Table \ref{table:surface_energies}), which is very similar to the previously reported 1.17 Jm$^2$ for $w$=3 \cite{Zhao2018bc}. Comparing this to the surface energies of the non-reconstructed (001) surfaces as well as the higher index non-polar (113) surface \cite{Ouhbi2018}, we find the NaO-terminated (001) surface to be the least stable followed by the reconstructed (001) and the TaO$_2$-terminated (001) surface, while the formally non-polar (113) surface (see Figure \ref{fig:slab_models}b) is the most stable (see Table \ref{table:surface_energies}). While the high energy of the NaO-terminated surface is not surprising, given the sequence of formally charged layers, the relatively low energy of the TaO$_2$-terminated surface is unexpected. We however believe it to be an artefact of our rather thin slabs as higher surface energies were reported for thicker slabs \cite{Zhao2018bc}. Taking this into considerations, the reconstructed (001) and the (113) surface both have low surface energies with a difference of only 0.08 J/m$^2$ and could both be exposed. As shown in Figure \ref{fig:slab_models}, these two surfaces differ in the coordination of their exposed Ta sites, which are square pyramidal (Ta$_{5c}$) coordinated on the reconstructed (001) surface but distorted tetrahedral coordinated (Ta$_{4c}$) on the (113) surface.

\begin{table}
	\caption{Surface energies of the non-reconstructed (both NaO- and TaO$_2$-terminated), reconstructed (001) and (113) surfaces of NaTaO$_3$.}
	\begin{tabular}{l l}
	\hline
	Surface & Surface energy (J/m$^2$)\\
	\hline 
	Reconstructed (001) & 1.20\\
	NaO-terminated (001) & 1.77\\
	TaO$_2$-terminated (001) & 1.17\\
	NaTaO-terminated (113) & 1.12 \\
	\hline 
	\end{tabular}
	\label{table:surface_energies}
\end{table}

\subsection{OER on clean surfaces}

\begin{figure}
	\centering
	\includegraphics{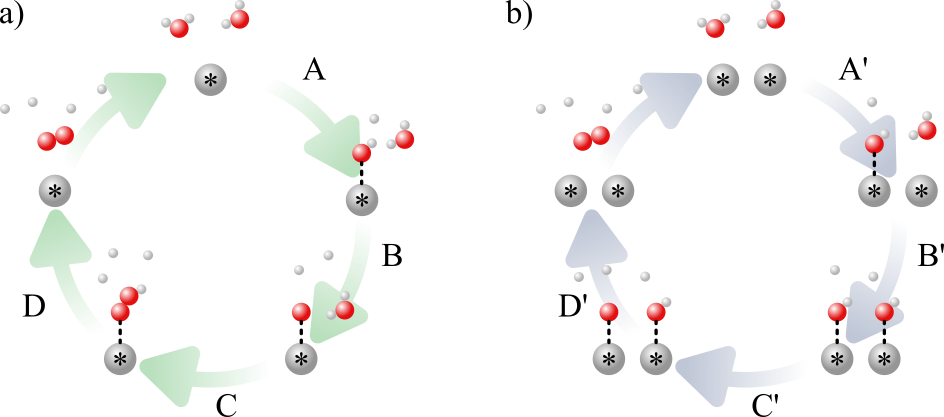}
	\caption{Schematic representation of a) the conventional and b) the coupling mechanism.}
	\label{fig:coupling_schematic}
\end{figure}

\begin{figure}
	\centering
	\includegraphics{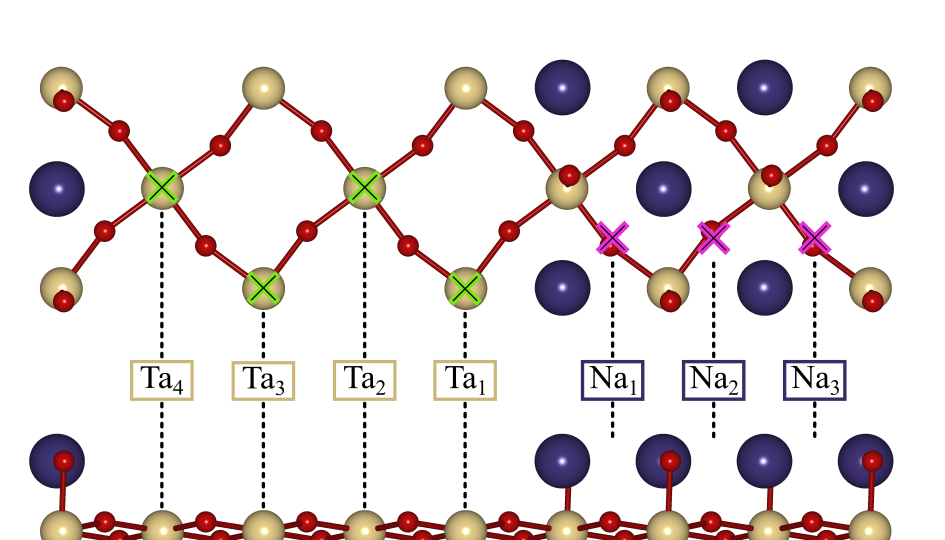}
	\caption{On-top Ta and bridging Na adsorption sites on the reconstructed (001) surface shown in a) top and b) side view. Only the surface layer is shown for clarity.}
	\label{fig:reconstructed_sites}
\end{figure}

To study the effect of the surface reconstruction on the OER activity, we computed free energy differences for the reaction steps on the reconstructed (001) surface. Among the reaction mechanisms reported in the literature, we adopt the conventional one \cite{Rossmeisl2007}, which is a succession of four proton-coupled electron transfer reaction steps as shown schematically in Figure \ref{fig:coupling_schematic}a). In the first step a water molecule is deprotonated and forms an adsorbed hydroxyl *OH, which is then deprotonated to form an adsorbed oxygen *O. In contact with water a further deprotonation step leads to an adsorbed hydroperoxyl group *OOH, which is released as O$_2$ after a final deprotonation step as shown by equations \ref{eq:A}-\ref{eq:D}:
\begin{eqnarray}
	\ce{2H2O + {*} &->[A]& {*}OH + H2O + (H+ + e^-)}\label{eq:A}\\
	\ce{&->[B]& {*}O + H2O + 2(H+ + e^-)}\label{eq:B}\\
	\ce{&->[C]& {*}OOH + 3(H+ + e^-)}\label{eq:C}\\
	\ce{&->[D]& O2 + 4(H+ + e^-)}\label{eq:D},
\end{eqnarray}
Here * is the active site on the surface, *OH, *O, *OOH are intermediates adsorbed on the active site, and A, B, C, D, are labels for the water oxidation steps. We consider as active sites all inequivalent Na$_i$ or Ta$_i$ sites (see Figure \ref{fig:reconstructed_sites}), where the former are top sites, while the latter are bridge sites between two adjacent Na atoms.

In Table \ref{table:binding_overpotentials} we report the adsorption energy of the OER adsorbates and the resulting overpotential for all inequivalent Ta and Na sites on the reconstructed (001) surface. The adsorption free energy with respect to water and hydrogen was computed as
\begin{equation}
	\Delta G_i = G_{\mathrm{slab}+i} - G_\mathrm{slab} - nG_\mathrm{H_2O} - mG_\mathrm{H_2},
\end{equation}
where the adsorbates are $i=\mathrm{OH, O, OOH}$, the free energies $G$ contain the DFT total energy as well as ZPE and entropy contributions and the coefficients $n$ and $m$ are chosen to provide the number of O and H atoms contained in the adsorbate. The differences between adsorption energies at different Ta sites are generally small. We see however a clear trend for less positive adsorption free energies (stronger binding) of all adsorbates at the step sites Ta$_1$ and Ta$_4$ compared to the terrace sites Ta$_2$ and Ta$_3$. Differences in binding geometry do not correlate with the binding strength (see supporting information Figure S1) and we hence associate the stronger binding at Ta$_1$ and Ta$_4$ with the interaction of the adsorbate with the Na ions at the step edge. For all Ta sites the overpotential-determining step (ODS) was found to be the formation of OOH (step C) and the lowest overpotential is about 0.70 V at the terrace sites Ta$_2$ and Ta$_3$. Since $\Delta G_\mathrm{O}$ - $\Delta G_\mathrm{OH}$ is larger for terrace than step sites, this is in good agreement with the universal scaling relations \cite{Man2011}.

\begin{table}
\caption{Binding energies ($\Delta G_i$) of the OER adsorbates ($i$ = OH, O, OOH) and the calculated overpotential ($\eta$) on Ta and Na sites of the reconstructed NaTaO$_3$ (001) surface.}
\begin{tabular}{c c c c c}
\hline
Site & $\Delta G_\mathrm{OH}$ (eV)& $\Delta G_\mathrm{O}$ (eV) & $\Delta G_\mathrm{OOH}$ (eV) & $\eta$ (V) \\
\hline 
Ta$_1$& 0.11 & 1.45 & 3.54 & 0.80\\
Ta$_2$& 0.23 & 1.72 & 3.65 & 0.70\\
Ta$_3$& 0.23 & 1.72 & 3.66 & 0.71\\
Ta$_4$& 0.07 & 1.46 & 3.47 & 0.78\\
\hline 
Na$_1$& 1.36 & 4.09 & 4.06 & 1.50\\
Na$_2$& 1.44 & 4.03 & 4.44 & 1.36\\
Na$_3$& 1.38 & 4.08 & 4.44 & 1.47\\
\hline 
\end{tabular}
\label{table:binding_overpotentials}
\end{table}

\begin{figure}
	\centering
	\includegraphics{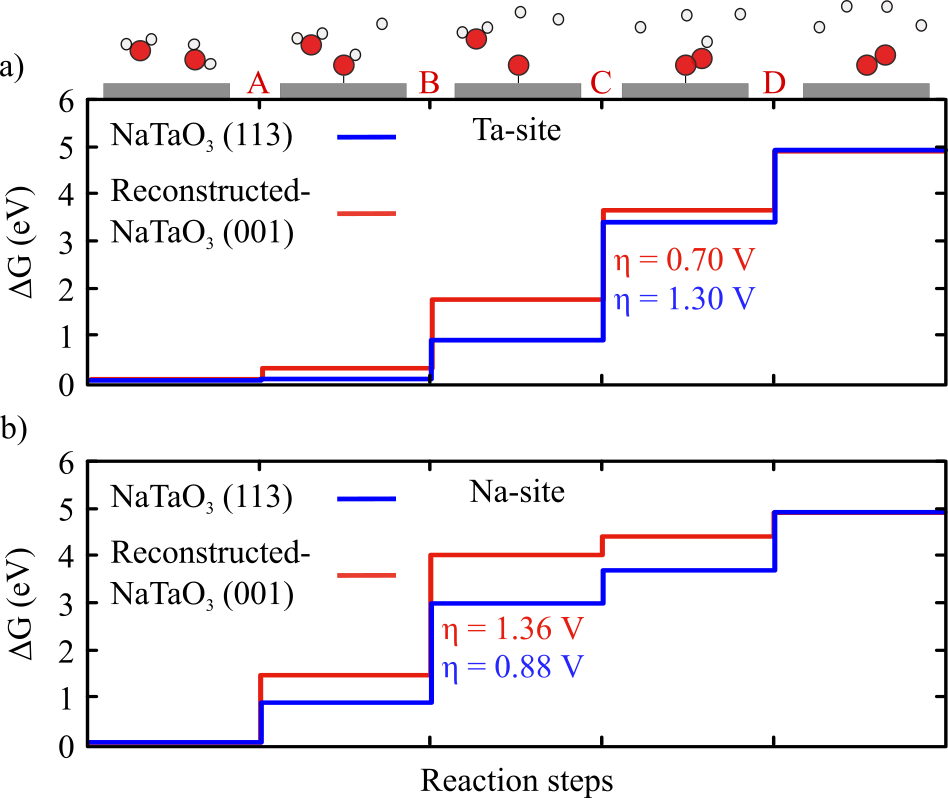}
	\caption{Free energy diagram at pH = 0 and U$_b$ = 0 V of the OER reaction steps at the most active a) Ta site and b) Na site of the reconstructed (001) surface. The profiles on the (113) surface \cite{Ouhbi2018} are shown for comparison.}
	\label{fig:free_energy_profile}
\end{figure}

Figure \ref{fig:free_energy_profile}a) shows the OER free energy profile at the Ta$_2$ site of the reconstructed (001) surface, comparing it with one for the Ta site on the (113) surface \cite{Ouhbi2018}. While step C is the ODS for both surfaces, the Ta$_2$ site of the reconstructed surface is more active for the OER ($\eta$ = 0.70 V) than its counterpart on the (113) surface ($\eta$ = 1.30 V). This is consistent with the adsorption strength of the O intermediate, which is less positive (stronger) on the (113) surface ($\Delta G_\mathrm{O}$= 0.86 eV) compared to the reconstructed (001) surface ($\Delta G_\mathrm{O}$ = 1.72 eV). The difference in activity is likely a result of the different coordination of the Ta site. For both surfaces we observe an upwards relaxation of the Ta site upon O adsorption, increasing the Ta--O bond length to an oxygen below the surface from 1.83 \AA\ to 2.52 \AA\ ($\Delta l$ = 0.69 \AA) for the reconstructed surface and from 1.90 \AA\ to 2.88 \AA\ ($\Delta l$ = 0.98 \AA) for the (113) surface. The 4-fold coordination on the (113) surface thus result in a higher structural flexibility and thus a stronger adsorption compared to the 5-fold coordinated Ta on the reconstructed (001) surface (see Figure \ref{fig:slab_models}).  Recently an overpotential of 1.43 V was reported for the non-reconstructed polar TaO$_2$ terminated (100) surface of NaTaO$_3$ with the ODS being step C \cite{Montoya2018}, which indicates that the reconstruction has a beneficial effect on the OER activity. 

On the NaO terrace, we find bridge adsorption sites to be most favoured (see Figure \ref{fig:reconstructed_sites}). As shown in Table \ref{table:binding_overpotentials} the difference in adsorption energies of the intermediates are fairly small for the three Na sites and result in a difference in overpotential of about 0.1 V, the ODS being the deprotonation of OH (step B) in all cases. The Na$_2$ site at the center of the terrace has a lower overpotential (1.36 V) than Na$_1$ and Na$_3$ sites more towards the edge of the terrace. Comparing with the (113) surface, which has the same ODS (step B), we find that the most active Na sites on the reconstructed (001) surface have a significantly higher overpotential (1.36 V) compared to the (113) surface (0.88 V) \cite{Ouhbi2018}.

\subsection{OER under operating conditions}

\begin{figure}
	\centering
	\includegraphics{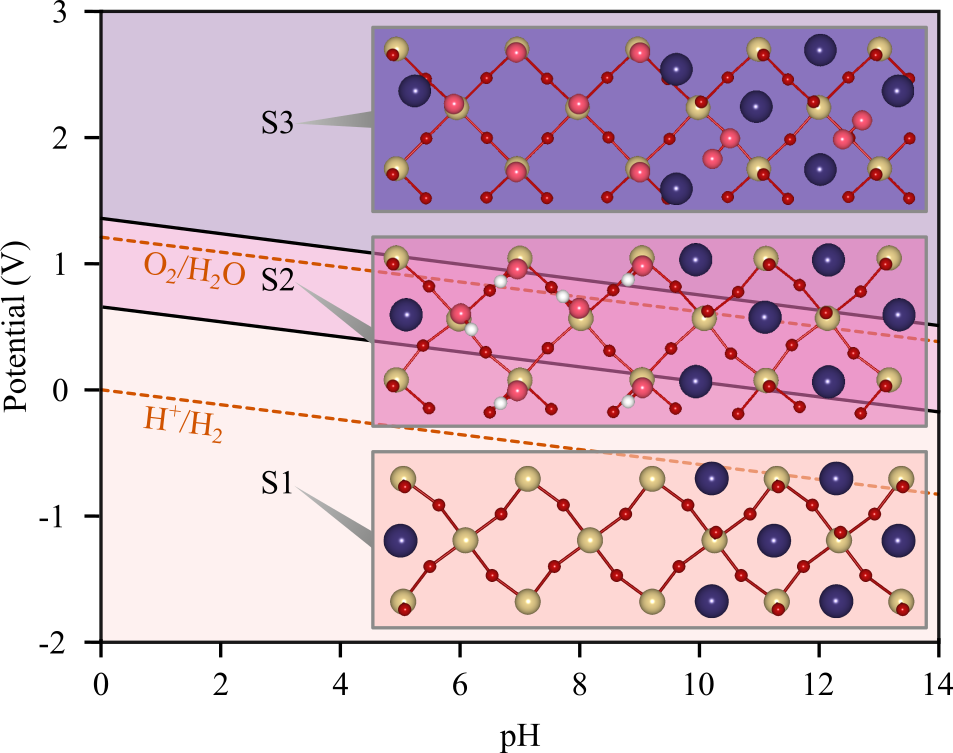}
	\caption{Computed Pourbaix diagram of the reconstructed (001) surface. Atomic structures show the different surface states, where pink and white atoms represent the adsorbate oxygen and hydrogen atoms respectively.}
	\label{fig:pourbaix}
\end{figure}

\begin{figure}
	\centering
	\includegraphics{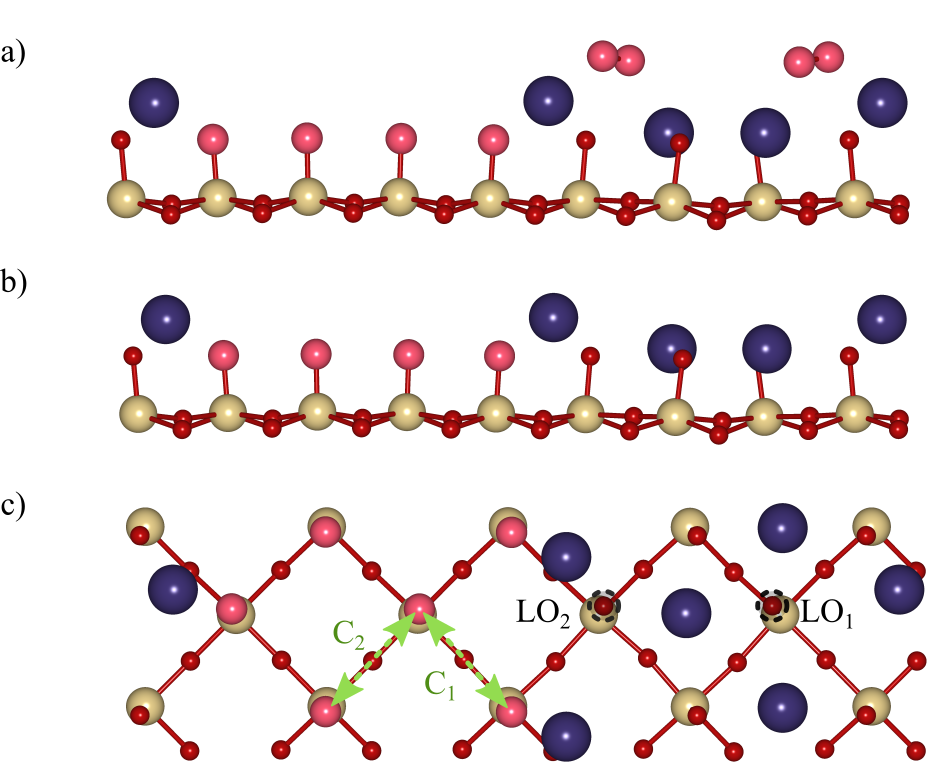}
	\caption{S3 state of the reconstructed-NaTaO$_3$ (001) surface a) before and after O$_2$ desorption shown in b) side and c) top view. The LO$_1$ and LO$_2$ sites marked in c) are the oxygen atoms participating in the lattice mechanism and green arrows show Ta pairs considered for the coupling mechanism.}
	\label{fig:state3}
\end{figure}

Next, we determine the surface adsorbate coverage on the reconstructed (001) surface as a function of the potential by computing Pourbaix diagrams as shown in Figure \ref{fig:pourbaix}. We explored separately covering the TaO$_2$ and NaO terraces with O and OH adsorbates at coverages ranging from 1/4 to 1 monolayer (ML) as well as configurations where both terraces are covered simultaneously. The clean surface is only exposed at potentials below 0.76 V at pH 0, before it becomes OH covered on the TaO$_2$ terrace and ultimately O covered on both terraces for potentials greater than 1.32 V. For the (113) surface \cite{Ouhbi2018}, we found a similar sequence of transitions, however at slightly lower potentials (0.66 and 0.99 V) and with an intermediate 1 ML O coverage on just the Ta sites. Interestingly, we observe for both surfaces a recombination of adjacent O adsorbed on Na sites (see state S3 in Figure \ref{fig:pourbaix} and Figure \ref{fig:state3}a).

Solar irradiation is expected to yield potentials above 1.32 V and we thus consider state S3 in Figure \ref{fig:pourbaix} as relevant under application conditions. Given that the O$_2$ formed on the NaO terrace is not bound to the surface, we remove it and consider a full O coverage only on the TaO$_2$ terrace as shown in side and top views in Figures \ref{fig:state3}b and \ref{fig:state3}c. We compute the OER on Ta sites, initially considering the conventional mechanism (Figure \ref{fig:coupling_schematic}a) starting with the formation of *OOH on a preexisting *O adsorbate (step C). We find an overpotential of 1.11 V at the step sites Ta$_1$ and Ta$_4$, while terrace sites Ta$_2$ and Ta$_3$ have an overpotential of 1.31 V. These overpotentials are higher than the ones reported above for the clean surface and the ODS changes from step C (OOH formation) on the clean surface to step B (O formation) on the S3 surface (see supporting information Figure S2 a). These results are in contrast to the usual effect of an increased adsorbate coverage leading to lower overpotentials \cite{Montoya2015a} that we also observed for the (113) surface (1.30 V $\rightarrow$ 0.92 V). A potential explanation for this effect is that the adsorbate coverage partially conceals the structural reconstruction, the O covered Ta terrace being intermediate between a TaO$_2$ and a NaO termination.

In addition to the conventional mechanism at Ta sites, we also consider an alternative mechanisms that proceeds by direct coupling of O adsorbates at adjacent sites (see Figure \ref{fig:coupling_schematic} b) \cite{Lodi1978, Norskov2004}. While this mechanism was not possible on the (113) surface \cite{Ouhbi2018} due to the large distance of 5.47 \AA\ between Ta sites, the (001) surface has shorter Ta-Ta distances of $\sim$ 3.87 \AA. We consider the coupling between the sites marked with C$_1$ and C$_2$ arrows in Figure \ref{fig:state3}c and find overpotentials of 0.88 V and 0.95 V respectively, the deprotonation of the first OH (step C') being the ODS in both cases (see supporting information Figure S2 b).

On the NaO terrace, we consider both the conventional mechanism and coupling of neighboring O adsorbates and find both mechanisms to have the same overpotential of 1.11 V with the formation of *OH as the ODS. The (113) surface on the other hand had a significantly lower overpotential of 0.79 V for the coupling mechanism \cite{Ouhbi2018}. In addition, we notice a tendency for OH and O adsorbates on the NaO terrace to migrate away from the bridge site and to bind with lattice oxygen. We did observe this neither on the NaTaO$_3$ (113) nor the SrO-terminated SrTaO$_2$N (001) surface \cite{Ouhbi2018}, indicating a particular behaviour of the reconstructed NaTaO$_3$ (001) surface. For this reason we investigate a lattice oxygen evolution reaction (LOER) mechanism \cite{Fabbri2018} at sites LO$_1$ and LO$_2$ marked in Figure \ref{fig:state3}c. We find overpotentials of 0.32 V and 1.09 V respectively, step D (recovery of the lattice oxygen by deprotonation of OH at the LO site) being the ODS for both sites (see supporting information Figure S2 c) and the OOH intermediate being unstable, transferring its H to a neighboring oxygen.

\begin{figure}
	\centering
	\includegraphics{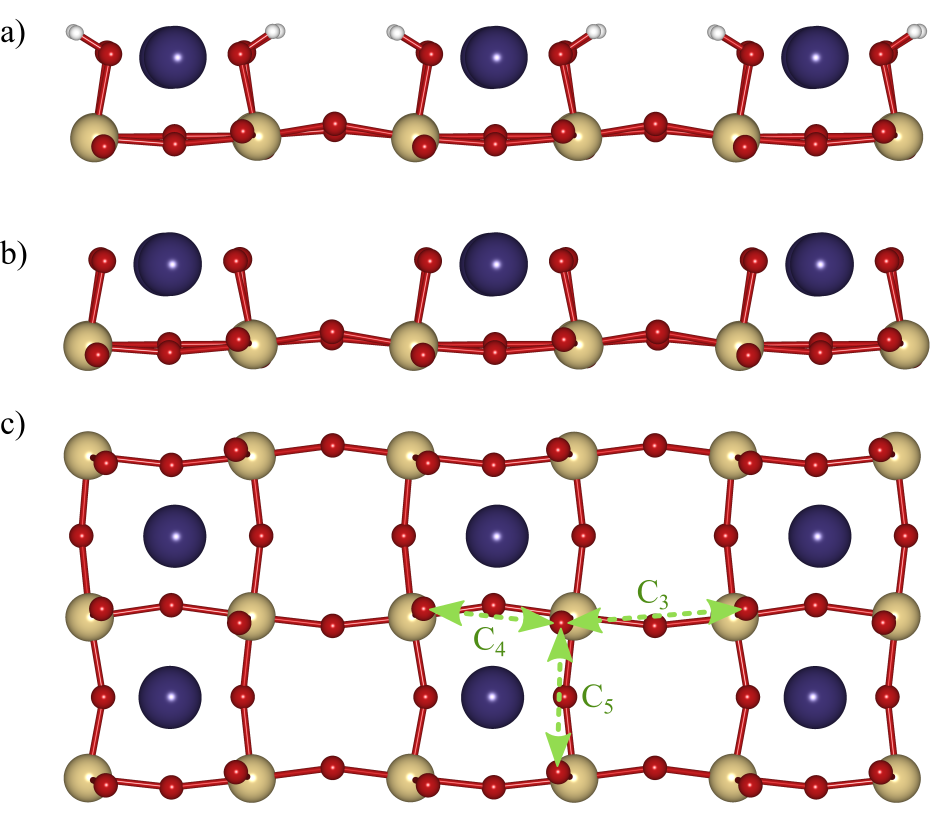}
	\caption{Surface reconstruction after exposure to water: a) side view at zero potential, b) and c) side view and top view under photocatalytic conditions respectively. Green arrows represent the various coupling pathways.}
	\label{fig:humid_structure}
\end{figure}

Above we discussed the OER on the reconstructed surface as it was observed under ultra-high vacuum (UHV) conditions \cite {Setvin2018}. However in the same study the authors report a second reconstruction with 2x1 periodicity that is based on hydroxylated steps running along the [100] direction and which forms after exposing the surface to water, which was confirmed in a later study \cite{Zhao2018bc}. We also consider the OER on this reconstruction (shown in Figure \ref{fig:humid_structure}a), which experimentally coexists with the one considered in the previous section \cite{Setvin2018}. We however assume that under photocatalytic conditions, the hydroxyls are deprotonated as shown in Figure \ref{fig:humid_structure}b) and c). Given the stepped structure of the surface, the conventional mechanism starting from an O at the step edge, can also be considered equivalent to the above LOER pathway on the NaO terrace. For this mechanism, we compute an overpotential of 0.64 V with step C (OOH formation) being the ODS. In analogy to what was shown above for the UHV reconstruction, the recombination of neighboring oxygen adsorbates represents an alternative pathway. We consider coupling of oxygen adsorbates from adjacent steps (pathway C$_3$), across a step (pathway C$_4$) and along the step edge (pathway C$_5$) as shown in Figure \ref{fig:humid_structure}c. We find the lowest overpotential of 0.52 V for pathway C$_3$, which does not involve interaction with Na atoms during coupling, while pathways C$_4$ and C$_5$ both have a larger overpotential of 0.63 V. We note however that these coupling pathways may be associated with additional kinetic barriers \cite{Rossmeisl2007}.

\section{Conclusions}

We have studied the OER activity on reconstructed (001) surfaces of NaTaO$_3$ by DFT calculations. On the clean UHV reconstructed surface we find the most active Ta and Na sites to have overpotentials of 0.70 and 1.36 V respectively. Compared to our previous results for NaTaO$_3$ (113) \cite{Ouhbi2018}, this implies that the Ta site is significantly more reactive on the reconstructed (001) surface, while the Na site is less reactive. Under photocatalytic conditions, the Ta sites are covered by O adsorbates, which leads to high overpotentials of $\sim$ 1.1 V. The adsorbate coverage however also results in a change in OER mechanism on the NaO terrace, where instead of the conventional mechanism, we now find a lattice oxygen evolution mechanism to be most active with a very small overpotential of 0.32 V. On an alternative surface reconstruction that was reported to be stable under exposure to water, we find a coupling mechanism between adjacent rows to be most active with an overpotential of 0.52 V. These results show a high sensitivity of the OER activity on the surface orientation and concomitant reconstructions of the same material. Moreover we show how different reconstructions of the same surface orientation can lead to dramatically different OER mechanisms and very different activities.

\begin{acknowledgement}

This work was funded by SNF Professorship Grant PP00P2\_157615. Calculations were performed on UBELIX (http://www.id.unibe.ch/hpc), the HPC cluster at the University of Bern.

\end{acknowledgement}


\bibliography{library}

\end{document}